\documentclass[12pt]{iopart}

\usepackage{comment}
\usepackage{graphicx}
\usepackage{float}
\usepackage{subfigure}
\usepackage{tabularx}
\usepackage{cite}

\usepackage{caption}
\usepackage{amsmath,amsfonts,amssymb}
\usepackage{iopams}
\usepackage{subfiles} 
\usepackage{url}
\usepackage{color}
\usepackage{xcolor}

\usepackage{tikz}
\usepackage{scalerel}
\usetikzlibrary{svg.path}
\definecolor{orcidlogocol}{HTML}{A6CE39}
\tikzset{
  orcidlogo/.pic={
    \fill[orcidlogocol] svg{M256,128c0,70.7-57.3,128-128,128C57.3,256,0,198.7,0,128C0,57.3,57.3,0,128,0C198.7,0,256,57.3,256,128z};
    \fill[white] svg{M86.3,186.2H70.9V79.1h15.4v48.4V186.2z}
                 svg{M108.9,79.1h41.6c39.6,0,57,28.3,57,53.6c0,27.5-21.5,53.6-56.8,53.6h-41.8V79.1z M124.3,172.4h24.5c34.9,0,42.9-26.5,42.9-39.7c0-21.5-13.7-39.7-43.7-39.7h-23.7V172.4z}
                 svg{M88.7,56.8c0,5.5-4.5,10.1-10.1,10.1c-5.6,0-10.1-4.6-10.1-10.1c0-5.6,4.5-10.1,10.1-10.1C84.2,46.7,88.7,51.3,88.7,56.8z};
  }
}
\newcommand\orcidicon[1]{\href{https://orcid.org/#1}{\mbox{\scalerel*{
\begin{tikzpicture}[yscale=-1,transform shape]
\pic{orcidlogo};
\end{tikzpicture}
}{|}}}}

\usepackage[hidelinks]{hyperref}

\begin{document}

\title[Optimal laser focusing for positron production in laser-electron scattering]{Optimal laser focusing for positron production in laser-electron scattering}

\author{
    Óscar Amaro \orcidicon{0000-0003-0615-0686} $^{1,2}$,
    Marija Vranic \orcidicon{0000-0003-3764-0645} $^{1,2}$
}

\address{$^1$GoLP/Instituto de Plasmas e Fus\~{a}o Nuclear, Instituto Superior T\'{e}cnico, Universidade de Lisboa, 1049-001 Lisbon, Portugal\\$^2$Authors to whom any correspondence should be addressed.}
\ead{\footnotesize
\href{mailto:oscar.amaro@tecnico.ulisboa.pt}{oscar.amaro@tecnico.ulisboa.pt}, \href{mailto:marija.vranic@tecnico.ulisboa.pt}{marija.vranic@tecnico.ulisboa.pt}
}\normalsize

\vspace{10pt}

\begin{abstract}

Laser-electron beam collisions that aim to generate electron-positron pairs require laser intensities $I\gtrsim10^{21}$ W/cm$^2$, which can be obtained by focusing a 1-PW optical laser to a spot smaller than 10 $\mu$m. Spatial synchronization is a challenge, because of the Poynting instability that can be a concern both for the interacting electron beam (if laser-generated) and the scattering laser. One strategy to overcome this problem is to use an electron beam coming from an accelerator (e.g. the planned E-320 experiment at FACET-II). Even using a stable accelerator beam, the plane wave approximation is too simplistic to describe the laser-electron scattering. This work extends analytical scaling laws for pair production, previously derived for the case of a plane wave and a short electron beam. We consider a focused laser beam colliding with electron beams of different shapes and sizes. The results take the spatial and temporal synchronization of the interaction into account, can be extended to arbitrary beam shapes, and prescribe the optimization strategies for near-future experiments.
\end{abstract}

\section{Introduction}\label{sc:intro}
In an intense electromagnetic background, charged particles obtain relativistic velocities and emit energetic photons. A fraction of these photons decays into electron-positron pairs, which can themselves be accelerated by the fields and radiate new photons \cite{Ritus1985,Erber66}. The repeated recurrence of photon emission and pair creation leads to the formation of the so-called QED cascades, where the number of particles in a plasma grows exponentially with time. 

Pair cascades are believed to affect plasma dynamics in extreme astrophysical environments, e.g. in pulsar magnetospheres and polar caps  \cite{astroplasma_Uzdensky_2014, Timokhin2010, astroplasma_Medin_2010}. It was recently proposed that one could re-create a comparable energy density in the laboratory using counter-propagating intense laser pulses \cite{Kirk2009}, which has prompted many scientists to study related configurations using kinetic particle-in-cell simulations \cite{Grismayer2016, Luo2018, Elkina2011, qu2020, Bulanov2013, Bell2008, Bulanov2006, Nerush2011, Mironov_2016, Lobet_2016, VRANIC201565, Duclous_2010, Gonoskov2015, Bashmakov2014, Gelfer2015, Vranic2016_trap, Gonoskov2017, Jirka2016, SeededQEDcascades2017, Kostyukov2016, Jirka2017}. As the required laser intensity ($I\sim 10^{24}~\mathrm{W/cm^2}$) is still beyond the extent of the current laser technology, there are many unknowns about the highly nonlinear dynamics associated with plasmas in extreme conditions. 

Before lasers become sufficiently intense to generate dense $e^+e^-$ pair plasmas from light, a head-on collision between a pulsed laser and a very energetic electron beam can allow us to generate dilute $e^+e^-$ beams by applying the currently available technology. The famous SLAC E-144 experiment has shown the onset of nonlinear Compton scattering and Breit-Wheeler pair production combining a $10^{18}~\mathrm{W/cm^2}$ laser with a $\sim 50~\mathrm{GeV}$  electron beam  \cite{slac1996,slac1997}, albeit with a very low positron yield.  Experiments are underway, expected to produce more pairs per shot \cite{Yakimenko2019}.
Research on stochastic effects in radiation reaction is also expected to benefit from the laser-electron scattering experiments \cite{DiPiazza2013,Vranic2016,Niel2018,Ridgers2017,Blackburn2015}, with new ways to infer the peak laser intensity at the interaction point \cite{tamburini2020,Blackburn2020ind} and probing the transition from the classical to the quantum-dominated laser-electron interaction. 
Two all-optical experiments have shown the electron slowdown due to radiation reaction  \cite{Cole2018, Poder2018}, but were not able to discriminate between different theoretical descriptions of radiation reaction. 
We anticipate the near-future facilities (e.g. ELI \cite{ELI}, Apollon \cite{Apollon}, CoReLS \cite{CoReLS}, FACET-II \cite{FACET-II,meuren2020seminal}, LUXE \cite{abramowicz2019letter}, EXCELS \cite{EXCELS}, ZEUS\cite{ZEUS}) are to probe the electron-positron pair production covering several different regimes of interaction. This manuscript focuses on head-on laser-electron scattering that maximizes the strength of the electric field in the electron rest frame. This is the first experiment planned in most of the aforementioned facilities, and we aim to to improve the current predictive capabilities for positron creation.

Due to the inherent non-linearity of the Breit-Wheeler pair production, there is no general roadmap on what would be an optimal strategy to obtain the highest possible positron yield using any given laser system. If the laser is assumed to be a plane wave (adequate when the laser is much wider than the interacting beam), the analytical predictions state that the best strategy would be to use the highest conceivable laser intensity. To achieve this goal, one is tempted to conclude that the laser should be focused on the smallest attainable focal spot. Our work shows that this strategy may not always be optimal, as  there is a trade-off between the high laser intensity and the size of the interaction volume. With a short focus, the highest intensity region becomes small both transversely and longitudinally, which can reduce the number of seed electrons that interact with the close-to-the-maximum intensity, as well as the duration of this interaction. Using tight focusing also increases the number of particles that are not temporally synchronized with the peak of the laser pulse at the focal plane (these electrons never get to interact with the peak laser intensity, regardless of the transverse position). Finally, the wavefront curvature can also change the effective angle of laser-particle interaction. 

Each of the mentioned factors affects the resulting number of positrons, hence a correct optimisation strategy would have to take all of them into account at the same time. This can be achieved by resorting to full-scale 3-dimensional particle-in-cell simulations, making sure enough statistics is used to represent the interacting electron beam with all its features, as well as high grid resolution in all spatial directions to correctly describe the laser dynamics. This approach requires a lot of computational resources (several million CPU-hours for each parameter set) and can be justified for the support of a specific on-going experiment where most parameters are not free. However, for future experiments, there are many possible choices. It would thus be practical to devise a simple and cost-effective way for their consideration. This would ensure that the best possible strategies are applied when constructing new laser facilities. 

This manuscript aims to simplify the pair production optimization in multi-variable parameter space. The goal is accomplished by extending the analytical scaling laws previously developed for electron interaction with a plane wave to more realistic geometries, taking into account the finite electron beam size and the laser focusing. Our method allows for predicting the number of positrons created in a laser-electron collision with a temporal (longitudinal) or a perpendicular offset.

This paper is structured as follows. In Section 2, we revisit the pair production in a plane wave. Section 3 covers different methods for calculating the overall positron yield. In Section 4, we give predictions for a special case where the electron beam is long and wide compared to the laser beam at the focus. Section 5 discusses a  thin electron beam, while Section 6 features a short electron beam. Finally, in Section 7 we show the optimisation of the positron number expected as a function of the available laser and electron beam parameters.

\section{Pair production in a plane wave}\label{sc:plane}

The simplest description of a laser pulse is a plane wave with a temporal envelope. Such a wave is fully described by the wavelength $\lambda$, a pulse duration $\tau$ (that defines the extension of the pulse's temporal envelope), and the normalized vector potential $a_0$, which relates to the intensity through $a_0 = 0.855 \sqrt{I[10^{18}~ \mathrm{W /cm^{2}}]} \lambda [\mu \mathrm{m}]$ (for linearly polarized lasers). 
As a relativistic electron interacts with the strong electromagnetic wavepacket, it emits high-energy photons that themselves interact with the laser field and can decay into electron-positron pairs. The process of electron-positron generation is mediated by a real photon, through the Breit-Wheeler mechanism \cite{breitwheeler1934}.

In the plane wave approximation, the total number of new pairs per interacting electron can be estimated if we know the electron energy $\gamma_0 m_e c^2$ (where  $\gamma_0$ is the initial electron Lorentz factor, $m_e$ the electron mass and $c$ is the speed of light), and the laser parameters (peak $a_0$, central wavelength $\lambda$ and pulse duration $\tau$, which is defined as the full width at half maximum of the laser intensity). The total number of pairs is then given by \cite{Blackburn2017}:

\begin{equation}
    N_+^{PW} (\gamma_0,a_0,\lambda,\tau) \simeq 3\sqrt{\dfrac{\pi}{2}} P_{\pm}(\omega_c) ~ \chi_{c,rr} ~ \dfrac{(\gamma_0 m c^2 - \hbar \omega_c)^2}{\hbar \gamma_0 mc^2} ~\dfrac{dN_\gamma}{d\omega}  \bigg|_{\omega=\omega_c}
    \label{eq:planewave}
\end{equation}

\noindent The first term $P_{\pm}(\omega_c)$ represents the probability of emitting a photon of frequency $\omega_c$; the second is the recoiled $\chi_{c,rr}$ which accounts for the radiation reaction on the beam electrons and the final term  $dN_\gamma/d\omega$ is the value of the emitted photon distribution at $\omega=\omega_c$. This approximation underestimates the number of low-energy photons, which does not significantly affect  the positron production calculation. According to this model, all positrons are generated from photons with a critical frequency $\omega_c$ and there is no feedback by the produced pairs on the photon spectra (in other words, there is only one generation of secondary particles).

\begin{figure}[ht]
\centering
    \includegraphics[width=200pt]{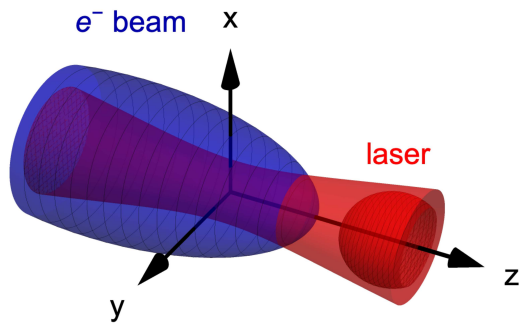}
  \caption{Illustration of an electron beam colliding with a Gaussian laser pulse.}
  \label{fig:pw}
\end{figure}

\section{Beyond plane wave}\label{sc:beyond}

Let us now consider a diffraction-limited laser pulse illustrated in figure \ref{fig:config}. The maximum laser intensity an individual particle within the electron beam interacts with depends on two geometrical factors: the transverse offset from the laser axis compared to the laser spotsize and the initial longitudinal position that affects the temporal synchronization of the interaction. In other words, while interacting with a Gaussian laser pulse, electrons far from the focus interact with a lower average (and peak) field, which must be taken into account. The electron encounters the peak of the laser pulse at time $t$ in an $(x,y,z)$ point of configuration space which defines the  maximum field felt by this particle. We can therefore assign an effective vector potential $a_{0,\textrm{eff}}(t, x, y, z)$ that corresponds to the maximum laser intensity the particle experiences during the interaction.

We define an equivalent distribution of beam particles according to the maximum intensity they interact with during the scattering. This intensity is identified through the maximum instantaneous vector potential associated with an individual beam particle as $a_{0, \text{eff}}$. In the case of a plane wave interaction there is no defocusing and particles always interact with the same intensity, regardless of where or when they overlap with the peak of the laser ($a_{0,\text{eff}}\equiv a_0$, and the equivalent distribution in this case would be a dirac Delta function).
For a more general case, by considering a corrected $a_{0, \text{eff}}$ for each particle, we can apply the equations already derived for a plane wave (equation \ref{eq:planewave}), and then integrate over the distribution function in  $a_{0, \text{eff}}$ to obtain the total yield of positrons in the laser-electron scattering. The integration can be performed by sampling the distribution function numerically, or performing an analytical integration over the configuration space in some special cases where this is possible (e.g. the authors in \cite{Blackburn2017} took this approach for a special case compared in section \ref{sc:short} with our result). 

Our approach is to first obtain a distribution of the interacting particles according to their $a_{0,\mathrm{eff}}$. For every bin in this distribution $dN(a_{0,\mathrm{eff}})/da_{0,\mathrm{eff}}$ one can calculate the contribution for pair production using $N_+^{PW} (\gamma_0,a_{0,\mathrm{eff}},\lambda, \tau)$. This method is cost-effective because it eliminates the need to perform multiple variable integration in the configuration space.

\begin{figure}[!h]
\centering
    \includegraphics[width=150pt]{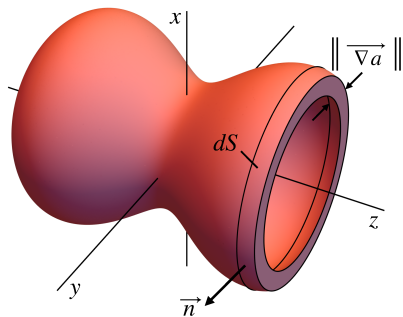}
  \caption{A volume element between two isosurfaces of the effective normalized vector potential $a_{0,\mathrm{eff}}$. The volume element contains all the  points where particles experience the peak $a_0$ within the interval $(a_{0,\mathrm{eff}},a_{0,\mathrm{eff}}+d a_{0,\mathrm{eff}})$.}
  \label{fig:schematic}
\end{figure}

\begin{figure}[!h]
\centering
    \includegraphics[width=440pt]{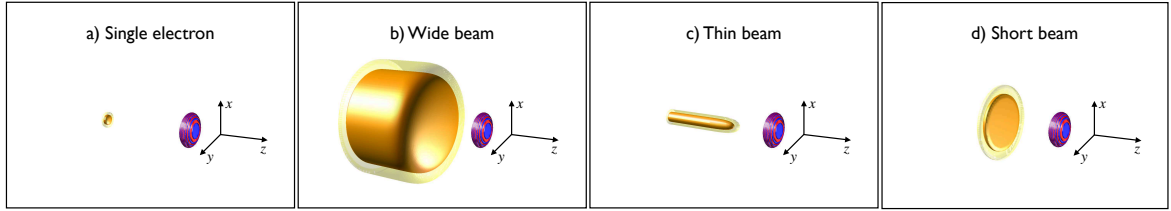}
  \caption{Scattering with nontrivial electron beam shapes. a) A single electron-laser interaction equivalent to electron colliding with a plane wave packet ($L \ll z_R,\allowbreak~R \ll W_0$). b) Interaction with a long and wide electron beam ($L \gtrsim z_R,\allowbreak~R \gg W_0$). c) Interaction with a pencil-like thin electron beam ($R \ll W_0$). d) Interaction with a short electron beam ( $L \ll z_R$). }
  \label{fig:config}
\end{figure}

The problem can be addressed using cylindrical coordinate system $(\rho,\phi,z)$, centered at the laser focus. For a Gaussian laser beam (in the paraxial approximation), the configuration space can be mapped according to the laser intensity isosurfaces shown in figure \ref{fig:schematic}, that do not depend on the coordinate $\phi$. For simplicity, let us first assume that the electron beam is a cylinder with a constant density $n_b$. Each particle meets the laser beam at a different point of space, and is assigned $a_{0,\text{eff}}(\rho,z)$, where $\rho$ and $z$ are its coordinates at the instant of time when it is synchronised with the peak of the laser. Performing a one-to-one mapping to the new coordinates of a flat-top relativistic beam in counter-propagation with the laser, the beam density in the new coordinates doubles and the length shrinks by two because the laser-electron crossing happens at twice the speed of light. The number of particles $dN_b(a_{0,\mathrm{eff}})$ with $a_{0,\mathrm{eff}}$ that falls in the interval $(a_{0,\mathrm{eff}},a_{0,\mathrm{eff}}+d a_{0,\mathrm{eff}})$ can then be estimated to be $dN_b(a_{0,\mathrm{eff}})/d a_{0,\mathrm{eff}} = 2n_b ~ dV/da_{0,\mathrm{eff}}$, where $dV$ is the volume between two adjacent isosurfaces associated with $a_{0,\mathrm{eff}}$ and $a_{0,\mathrm{eff}}+d a_{0,\mathrm{eff}}$. Due to the geometry of the problem, this expression can be transformed to the following:

\begin{equation}
    \frac{dN_b(a_{0,\mathrm{eff}})}{da_{0,\text{eff}}} = \int_S \frac{2n_b \ dS}{||\nabla a_{0,\text{eff}}||}
    \label{eq:histogram}
\end{equation}
where the surface element $dS = \rho \sqrt{d\rho^2+dz^2} ~d\phi = \rho \sqrt{1+(\partial\rho/\partial z)^2} ~dz ~d\phi$ is calculated at the isosurface that is by definition perpendicular to the gradient of the vector potential given by $||\nabla a_{0,\text{eff}}||= \sqrt{\left(\partial a_{0,\text{eff}}/\partial \rho\right)^2+\left(\partial a_{0,\text{eff}}/\partial z\right)^2}$.

Letting beam plasma density vary in space $n_b(\vec{r})$ allows considering cases of short or long, wide or narrow beams, including non-ideal spatio-temporal synchronisation with the laser (as discussed later for cases illustrated in figure \ref{fig:config}). It is worth noting that even a point-particle interaction with a Gaussian beam is not equivalent with a plane wave approximation unless the  particle is in perfect temporal synchronization with the laser pulse. 

Once the particle distribution in equation \eqref{eq:histogram} is calculated, we can extract field moments $\langle a_{0,\text{eff}} {}^k \rangle = \int a_{0,\text{eff}} {}^k \ dN/da_{0,\text{eff}} \ da_{0,\text{eff}}$, which can for example be used to calculate the average laser intensity \cite{Blackburn2020ind}.

\section{Wide beam}\label{sc:long}

As a first application of the ideas presented in the last section, let us consider the case of the scattering between a focused Gaussian laser pulse and a wide electron beam. The spatio-temporal intensity distribution of a Gaussian laser is characterized by the peak vector potential $a_0$, the laser wavelength $\lambda$ and the Rayleigh-range $z_R=\pi W_0^2/\lambda$, where $W_0$ is the transverse spot size. The effective vector potential has the following spatial dependence $a_{0,\textrm{eff}} = \left(a_0/\sqrt{1+(z/z_R)^2} \right) \exp \left( -(\rho^2/W_0^2)
/(1+(z/z_R)^2) \right)$,
where $z$ is the distance from the focal plane and $\rho$ is the distance from the laser propagation axis. 
Our definition of "a wide beam" is that the beam radius is much larger than the laser focal spot $W_0$. The gradient of $a_{0,\text{eff}}$ can be written as $||\nabla a_{0,\text{eff}}||= |\partial a_{0,\text{eff}}/\partial \rho | ~ \sqrt{1+(\partial\rho/\partial z)^2} $, where $|\partial a_{0,\text{eff}}/\partial \rho |=2\rho~ a_{0,\text{eff}}/(W_0^2(1+(z/z_R)^2)) $. This simplifies the particle distribution in $a_{0,\text{eff}}$  according to equation \eqref{eq:histogram}: 
\begin{equation}
        \frac{dN_b(a_{0,\mathrm{eff}})}{da_{0,\text{eff}}} = \frac{2\pi~n_b W_0^2}{a_{0,\mathrm{eff}}} \int_{z_\mathrm{min}}^{z_\mathrm{max}}1+\left(\frac{z}{z_R}\right)^2~dz
    \label{eq:integral}
\end{equation}
where the limits of integration in $z$ direction will depend on the beam length, and its temporal synchronization with the laser pulse. If the entire isosurface associated with a specific $a_{0,\text{eff}}$ is covered with interacting particles, $z_\mathrm{max}=-z_\mathrm{min}=z_R \sqrt{(a_0/a_{0,\text{eff}})^2-1}$. Otherwise, a portion of the volume associated with a specific laser intensity may be empty due to the finite beam length and temporal synchronization. The interaction limits imposed by the beam are $z^*_{\textrm{min}} = \Delta_\parallel -L/4$ and $z^*_{\textrm{max}} = \Delta_\parallel +L/4$, where $\Delta_\parallel$ is the longitudinal displacement of the electron beam centre from the focal plane when the laser is at the focus.  
For every $a_{0,\text{eff}}$, one has to evaluate what the appropriate integration limits are on each side, and there is a transition in the distribution function at values of $a_{0,\text{eff}}$ that corresponds to the beam edge on axis (as shown by examples in figure \ref{sc:long}a). The distribution function for a centered, wide, flat-top electron beam is given explicitly in Table \ref{tb:dist}.

We now illustrate the obtained particle distributions according to the effective laser intensity on an example. The SFQED experiment \cite{FACET-II} will study pair-production using a $0.61$ J laser pulse ($a_0=7.3$, $\lambda=0.8~ \mu \textrm{m}$, $W_0=3 ~\mu \textrm{m}$, $\tau = 35~ \textrm{fs}$) and a $13$ GeV, $2~\textrm{nC}$ electron beam. The electron beam follows a non-symmetric Gaussian density distribution transversely with $\sigma_x=24.4~\mu \textrm{m}$, $ \sigma_y=29.6 ~\mu \textrm{m}$, and has a $\sim 250~\mu\mathrm{m}$ long flat-top longitudinal profile.

\begin{figure}[!h]
\centering
    \includegraphics[width=420pt]{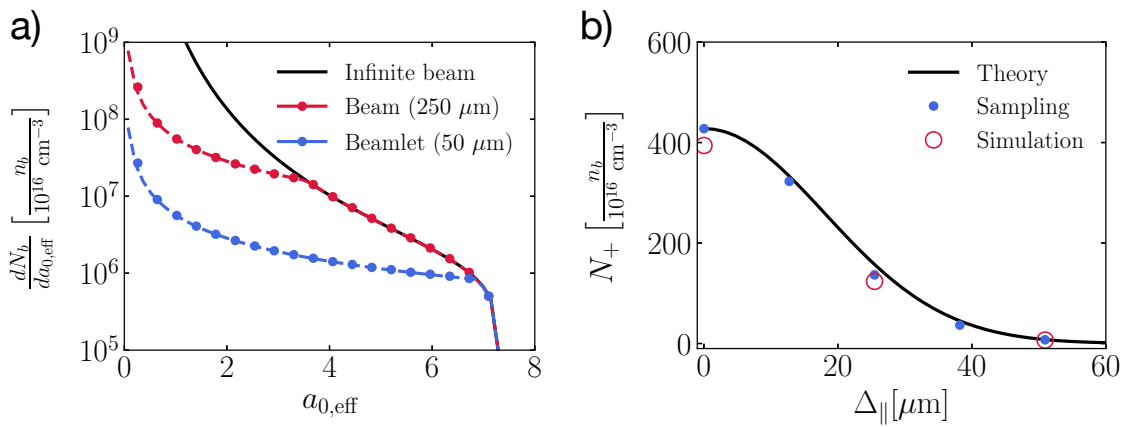}
  \caption{Wide beam, with a flat-top longitudinal envelope. a) Particle distribution according to the effective laser vector potential $a_{0,\mathrm{eff}}$. Dashed lines are analytical expressions, circles are from sampling and full line corresponds to the limiting case of $L\rightarrow \infty$. b) Positron yield as a function of the longitudinal displacement $\Delta_\parallel$ for one beamlet. The laser parameters are $a_0=7.3$, $\lambda=0.8~ \mu \textrm{m}$,  $\tau=31 ~\textrm{fs}$ and  $W_0=3~\mu \textrm{m}$. The beam energy is $E_0=13~ \textrm{GeV}$, transverse width $\sigma_x = 24.4 ~\mu \textrm{m}$ and $\sigma_y =~ 29.6~ \mu \textrm{m}$. The beam length is $L=50.9~ \mu \textrm{m}$ for each beamlet, while the entire beam contains $Q=2~\textrm{nC}$ and is $250~\mu\mathrm{m}$-long.}
  \label{fig:3D}
\end{figure}

To save computing time, we performed 3D PIC simulations of this interaction using OSIRIS \cite{Fonseca2002} by dividing the long beam in five equal beamlets each $50.9 ~\mu \textrm{m}$ long. These beamlets have different temporal synchronization (they encounter the laser peak at different distances from the focus).
The 3D simulations are performed with a box size of $98 ~\mu \textrm{m} \times 25~ \mu \textrm{m} \times 25 ~\mu \textrm{m}$, resolved with $3840 \times 400 \times 400$ cells. OSIRIS PIC results (red empty circles in Fig. \ref{fig:3D}(b)) are compared with the analytical predictions based on above intensity distribution functions and a numerically sampled beam.

For the analytical calculations and numerical sampling, we assumed the beam has a uniform density equal to the central density of the electron beam $n_b=10^{16}~\mathrm{cm^{-3}}$. This is justified by $\sigma_x\gg W_0$ and $\sigma_y\gg W_0$, and the highest intensity portion interacts nearly exclusively with the maximum density of the beam. The analytical calculation is therefore coherent with the simulation results, as confirmed by the comparison in figure 4b.

\section{Thin beam}\label{sc:thin}

Let us now consider the case of the scattering between a focused Gaussian laser and a long $L\gg z_R$  but thin $R\ll W_0$ electron beam, where the effective laser intensity can be lower due to a longitudinal offset $\Delta_\parallel$. Here, the problem becomes one-dimensional as the number of particles in the volume associated with a specific value of intensity is given by $dN_b=2n_b ~dV=4n_b ~S_\perp ~dz=4 (N_b/L) ~dz $, provided that the beam is long enough to interact with the two portions of the same laser intensity before and after the focus (which is always the case if the beam is temporally centered - otherwise, if the interaction happens only on one side, the total number of particles should be divided by two). The effective laser intensity depends only on $z$ through $a_{0,\text{eff}}=a_0/\sqrt{1+(z/z_R)^2}$ and the distribution can simply be calculated as

\begin{equation}
     \frac{dN_b}{da_{0,\text{eff}}}=\frac{4 N_b}{L}\frac{dz}{da_{0,\text{eff}}}.
\end{equation}

Just as in the previous section, the explicit distribution function in $a_{0,\text{eff}}$ for a centered beam is given in Table \ref{tb:dist}.

\begin{figure}[!h]
\centering
    \includegraphics[width=420pt]{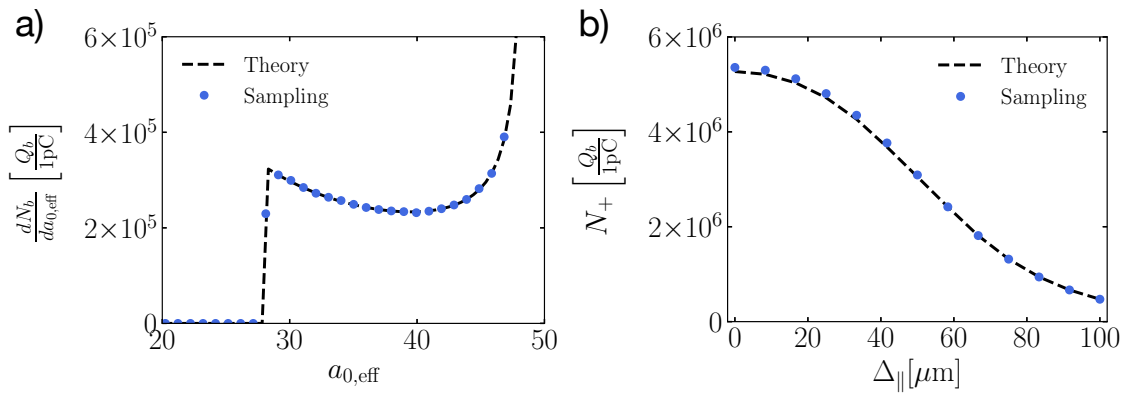}
    \caption{Thin beam. a) Particle distribution according to the effective vector potential for $\Delta_\parallel = 0$. b) Positron yield as a function of temporal synchronisation. The beam energy is $E_0=13 ~\textrm{GeV}$, charge is $Q_b=~1~\textrm{pC}$ and length $L=200~\mu \textrm{m}$. The laser parameters are $a_0=48.4,\allowbreak ~\lambda=0.8 ~\mu \textrm{m},~ \tau=35~ \textrm{fs}, ~W_0=3~\mu \textrm{m}$ and $\Delta_\parallel$ represents the longitudinal displacement of the beam centre when the laser is at the focus.}
    \label{fig:3DL}
\end{figure}

\section{Short beam}\label{sc:short}

Let us now consider a short $L\ll z_R$ beam with a Gaussian density profile $n_b = n_0 \exp \left( -((x-\Delta_\perp)^2+y^2)/R^2 \right)$ where the peak density is given by $n_0 = N_b/(\pi R^2L)$, $x=\rho \cos \phi$ and $y=\rho \sin \phi$. We assume  a longitudinally synchronized beam ($\Delta_\parallel=0$), with an allowed transverse displacement $\Delta_\perp$ between the beam centre and the laser propagation axis. The electrons therefore interact with the laser peak at $z=0$ and the field structure reduces to $a_{0,\text{eff}} = a_0 \exp \left( -\rho^2/W_0^2 \right)$. 
As the manifolds of constant $a_{0,\text{eff}}$ are now concentric rings, the volume element associated with a specific value of $a_{0,\text{eff}}$ is given by $dV = L~\rho~ d\rho ~d\phi/2$, the surface element of an isosurface is  $dS = L~\rho ~d\phi/2$, and the field gradient is given by $\nabla a_{0,\text{eff}} = \partial a_{0,\text{eff}}/\partial \rho \ \hat{\rho}$, with $\partial a_{0,\text{eff}}/\partial \rho = -2\rho \ a_{0,\text{eff}}/W_0^2$. 
We can now apply the equation \eqref{eq:histogram} to obtain the particle distribution function
$dN_b(a_{0,\mathrm{eff}})/d a_{0,\mathrm{eff}}=\int L ~n_b ~\rho~d\phi/||\nabla a_{0,\mathrm{eff}}|| $. This gives
\begin{equation}
     \frac{dN_b}{da_{0,\text{eff}}} = \frac{L n_0 W_0^2}{2 a_{0,\mathrm{eff}}} ~e^{-\rho^2/R^2}~e^{-\Delta_\perp^2/R^2}~\int_0^{2\pi}e^{2\rho \Delta_\perp \cos \phi/R^2}d\phi 
    \label{eq:3DT_analytical_integral}
\end{equation}
where  $-\rho^2 = W_0^2 \log (a_{0,\text{eff}}/a_0)$ and the integration result can be expressed through the modified Bessel function of the first kind $I_0(t)=(1/\pi)\int_0^\pi \exp(t\cos \phi)d\phi$. The obtained particle distribution function is given in Table \ref{tb:dist} and can be numerically integrated applying  $N_+^{PW}$ to every bin of the histogram to obtain the total positron yield from the laser-electron beam interaction. 
The calculations can be extended to a temporally non-synchronized interaction by replacing $a_0$ and $W_0$ with $a_0^*=a_0(z)=a_0/ \sqrt{1+(z/z_R)^2}$ and $W_0^*=W_0(z)=W_0 \sqrt{1+(z/z_R)^2}$.

\begin{figure}[!h]
\centering
    \includegraphics[width=420pt]{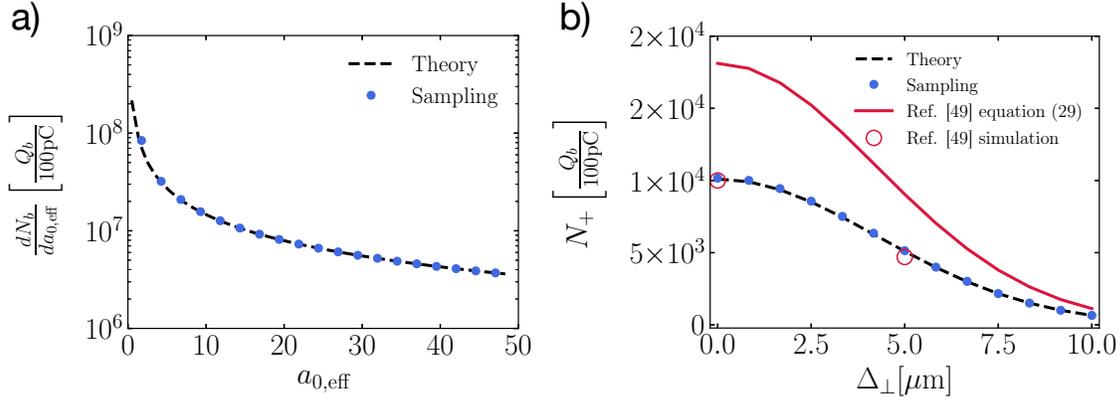}
    \caption{Short beam. a) Particle distribution according to the effective vector potential for a transversely aligned beam ($\Delta_\bot =0$). b) Positron yield as a function of transverse beam displacement from the laser propagation axis $\Delta_\bot$. The electron beam energy is $E_0=2~ \textrm{GeV}$, charge is $ Q_b=100~\textrm{pC}$ and Gaussian radius $R = 6~ \mu \textrm{m}$ in all spatial directions. The laser parameters are $a_0=48.4,\allowbreak \lambda=0.8 ~ \mu \textrm{m}, \tau=30 ~\textrm{fs}$ and $ W_0=~2\mu \textrm{m}$.}
    \label{fig:3DT}
\end{figure}

In Ref. \cite{Blackburn2017}, the authors consider a spherically-symmetric Gaussian beam profile with a radius $R=6~\mu \textrm{m}$, the laser spotsize $W_0 = 2~ \mu \textrm{m}$ and the Rayleigh range $z_R = 15.7~ \mu \textrm{m}$. As the Rayleigh range is much higher than the beam length ($z_R \gg R$), one can consider the beam short, just like in our calculations. They have obtained an approximate expression for the expected value of the number of positrons, which correctly predicts the order of magnitude, but has a factor of two difference compared with the simulation results in Ref. \cite{Blackburn2017}. We have applied our distribution functions to calculate the expected number of positrons in this case, and we have obtained the correct result that is consistent with the simulation data of Ref. \cite{Blackburn2017}. Detailed comparisons are shown in figure \ref{fig:3DT}.

\begin{table}[h!]

{\setlength{\extrarowheight}{10pt}
\begin{tabularx}{\textwidth}{l l} 
\hline
Setup & Particle distribution for temporally centered beams \\[5pt]
\hline\hline \\[-15pt]
Wide beam & 
    $\dfrac{dN_b}{da_{0,\text{eff}}} = \begin{cases}
        \dfrac{4\pi ~n_b ~W_0^2 ~z_R}{ a_{0,\text{eff}}} \dfrac{\sqrt{a_0^2-a_{0,\text{eff}}^2}}{3a_{0,\text{eff}} } \left( 2+\left(\dfrac{a_0}{a_{0,\text{eff}}}\right)^2 \right), a_{0,\text{eff}} \geqslant a_{z} \\
        \dfrac{4 \pi ~n_b ~W_0^2 ~z_R}{a_{0,\text{eff}}} \dfrac{L}{4z_R} \left(1+\left(\dfrac{L}{4z_R}\right)^2\right), a_{0,\text{eff}}<a_{z}
    \end{cases}$ \\[45pt]
 Thin beam & $\dfrac{dN_b}{da_{0,\text{eff}}} = \begin{cases}
        \dfrac{4N_b z_R}{L}\dfrac{a_0^2}{a_{0,\text{eff}}^2} \dfrac{1}{\sqrt{a_0^2-a_{0,\text{eff}}^2}}, a_{0,\text{eff}} \geqslant a_{z}\\[-5pt]
        0, a_{0,\text{eff}}<a_{z}
    \end{cases}$ \\[35pt]
 Short beam &
 $\dfrac{dN_b}{da_{0,\text{eff}}} = \dfrac{N_b~ W_0^2}{R^2 ~a_{0,\mathrm{eff}}} \left(\dfrac{a_{0,\text{eff}}}{a_0}\right)^{(W_0/R)^2}e^{-\Delta_\perp^2/R^2} ~I_0\left(\dfrac{2\Delta_\perp W_0}{R^2} \sqrt{\log \left(\dfrac{a_0}{a_{0,\text{eff}}}\right)}\right)$ \\[25pt]
 \hline
 \end{tabularx}
 }

 \caption{Particle distributions according to the effective vector potential for different beam geometries. Here, $a_{z} \equiv a_0/\sqrt{1+(L/4z_R)^2}
$, $N_b$ represents the total number of particles in the beam, $n_b$ is the beam density, $R$ and $L$ are the beam radius and length respectively. The laser spot size is $W_0$, $z_R\equiv \pi W_0^2/\lambda$ is the Rayleigh length, and $ \Delta_\bot$ is the perpendicular displacement of the beam centre from the laser propagation axis.}
 \label{tb:dist} 
\end{table}

\section{Optimal focusing strategy to obtain maximum positron yield}\label{sc:opt}

This section covers the optimal focusing strategy for a wide range of laser parameters (in particular as a function of total energy content and pulse duration), as well as different electron beam energies. We assume that the electron beam is 200 $\mu$m long (flat-top longitudinal envelope), and has a Gaussian transverse shape. The electron beam is spatio-temporally synchronized with the laser (i.e. the centre of the beam interacts with the laser peak at the focal plane, and they share the propagation axis). The transverse beam profile is Gaussian with $  \sigma_x = ~24.4 ~\mu \textrm{m}$ and $ \sigma_y = 29.6~ \mu \textrm{m}$. The chosen on-axis beam density $n_{b}=10^{16}~\mathrm{cm}^{-3}$ corresponds to the total beam charge of $Q_b=2~\textrm{nC}$. The results can be scaled to other values for the central beam density by introducing a factor $n_b/10^{16}~\mathrm{cm}^{-3}$.
A specific laser system has a fixed total energy content, which for a Gaussian transverse profile is approximately given by  $\varepsilon[\textrm{J}] \sim 2.1 \times 10^{-5}~ ~a_0^2~ (W_0/\lambda)^2 ~\tau [\textrm{fs}]$. The laser intensity (proportional to $a_0^2$) is therefore reversely proportional to the square of the spot size $W_0$. As the number of pairs produced per interacting electron $N_+^{PW}$ is a monotonously rising function of the effective $a_0$, and the number of seed electrons that would experience the high intensity is proportional to the size of the interaction volume $\sim W_0^2 z_R$, to obtain the highest possible number of positrons, one should strike the right balance between a high value of $a_0$ and a large $W_0$. In other words, there is a trade-off between using a short focal length to obtain the highest conceivable laser intensity, and having a wider interaction volume where more seed electrons participate in the interaction.

What follows is a calculation of the optimal focal spot and the corresponding pair yield for lasers with energy below $1~\mathrm{kJ}$ and relativistic particle beams with energies lower or equal to $20~\mathrm{GeV}$. 
These values include what will soon be available in several experimental facilities (e.g. SLAC \cite{FACET-II}, HiBEF \cite{abramowicz2021conceptual} or ELI \cite{ELI}).

\begin{figure}[!h]
\centering
    \includegraphics[width=420pt]{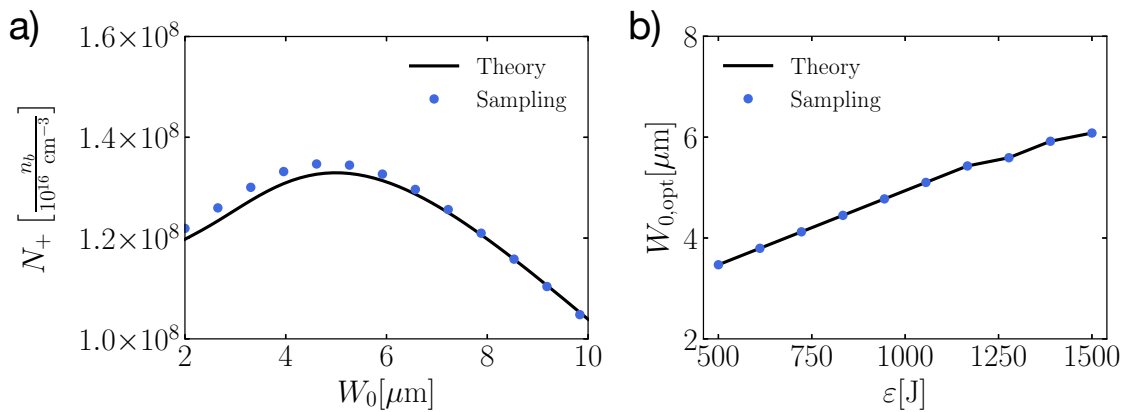}
    \caption{a) Positron yield as a function of the laser spot size keeping the total energy contained within the laser pulse constant at $\varepsilon = 1~\mathrm{kJ}$.  b) Optimal spot size for different total laser pulse energies. Beam parameters are $E_0=10~\mathrm{GeV},\allowbreak L=200~\mu\mathrm{m},\allowbreak \sigma_x = 24.4 ~\mu\mathrm{m},\allowbreak \sigma_y = 29.6 ~\mu\mathrm{m},\allowbreak n_b = 10^{16}~\mathrm{cm}^{-3}$, and we consider $\tau=150~\mathrm{fs}$  with $\lambda=0.8~\mu \mathrm{m}$. }
    \label{fig:opt}
\end{figure}

For each combination of the electron beam energy and the laser total energy content, we apply the analytical expression  (see Table \ref{tb:dist}) to calculate the effective $a_0$ distribution of the interacting particles. Then, we integrate the results numerically to find the optimal spotsize and maximum positron yield for this set of parameters (as illustrated in figure \ref{fig:opt}).

\begin{figure}[!h]
\centering
    \includegraphics[width=420pt]{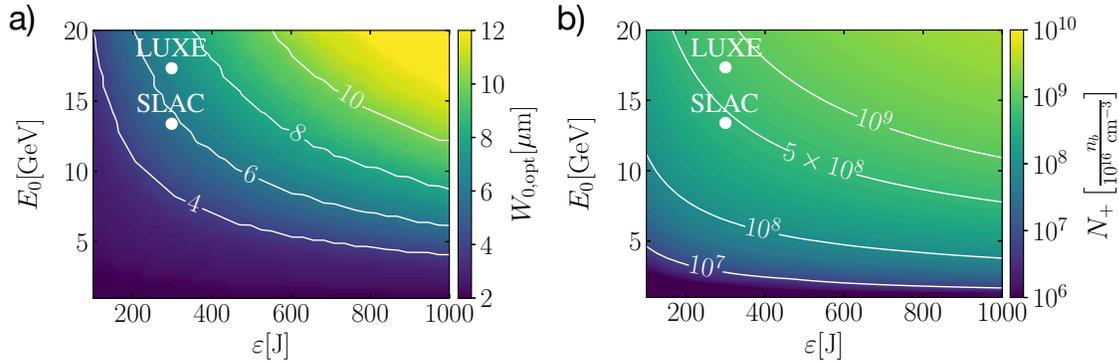}
    \caption{Optization study for lasers pulses of fixed duration ($\tau=35~ \textrm{fs}$). a) Optimal laser spotsize for a head-on scattering as a function of total pulse energy and the electron energy.  b) Positron yield achieved using the optimal spotsize. The laser wavelength is $\lambda=0.8~ \mu \textrm{m}$; the electron beam is $ L=200~ \mu \textrm{m}$ long (flat-top longitudinal profile) }
    \label{fig:TFg0}
\end{figure}

Figure \ref{fig:TFg0} summarizes  the optimization results, keeping the laser duration constant at $35~\mathrm{fs}$. For 10 GeV electrons and a 1 kJ laser, a maximum number of pairs is $~10^9$, which is obtained using $W_0>8~\mathrm{\mu m}$. The FACET-II 13 GeV electron beam at SLAC could generate $4\times10^8$ pairs/shot if paired with a  $300~\mathrm{J}$ laser focused to $W_0=5.7~\mu\mathrm{m}$. The LUXE $17.5~\mathrm{GeV}$ beam with the same laser parameters could produce $7\times10^8$ pairs per shot, using a slightly larger $W_0=6.8~\mathrm{\mu m}$.

\begin{figure}[!h]
\centering
    \includegraphics[width=420pt]{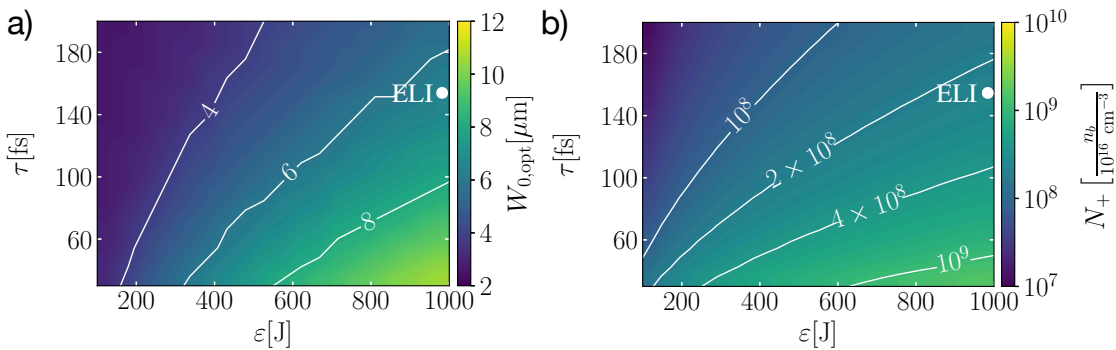}
    \caption{Optimization study for fixed electron beam parameters (here, electron energy is 13 GeV, corresponding to an accelerator beam available for example at SLAC).  a) Optimal laser spotsize for a head-on scattering as a function of pulse energy and duration. b) Positron yield achieved using the optimal spotsize. Other laser and electron parameters are the same as in figure \ref{fig:TFg0}}
    \label{fig:TFtau}
\end{figure}

Similarly, the fig \ref{fig:TFtau} shows how to obtain optimal results as a function of the laser energy and the laser pulse duration, keeping the initial electron beam energy constant at $13~\mathrm{GeV}$. 
This allows to estimate the positron yield at ELI Beamlines, where L4 laser specifications are at $1.5~\mathrm{kJ}$ with $150~\mathrm{fs}$ duration. If we assume a third of the laser energy is used to accelerate electrons, $1~\mathrm{kJ}$ is available for the scattering, which can produce $2.4\times10^{8}~(n_b/10^{16}~\mathrm{cm^{-3}})$ pairs per shot using $W_0=6.2~\mathrm{\mu m}$.

The results presented in this section best correspond to the case of a wide and long electron beam, which corresponds to a beam from a conventional accelerator. For LWFA beams, the "thin" beam may be a more adequate description for the laser-electron interaction in some cases, and we can perform a similar optimization using the thin beam effective intensity distribution functions.

\section{Conclusions}\label{sc:conclusions}

The methods outlined in this work allow to make predictions regarding the pair production in laser-electron scattering taking into account the 3D focusing geometry, spatio-temporal synchronization and the realistic beam shape and size. This opens a possibility for fast parameter optimization, using analytical calculations directly or combining them with a simple numerical integration. The approach is faster than using full-scale Monte Carlo or PIC-QED calculations, and the results can be obtained on a single CPU. Apart from saving computing resources, the ideas from the present study can be applied for real-time optimization and data analysis in experiments.        

Besides optimising the positron yield considered in this manuscript, the equivalent intensity distributions are going to be useful to calculate the asymptotic energy spread
\cite{DiPiazzaStochastic,GreenHarvey2014,Vranic2016,Ridgers2017, Niel2018} and divergence of the interacting electron beams
\cite{Yan2017, Mackenroth2019, tamburini2020, Hu2020}, which are also imprinted on the emitted photon beams in the hard x-ray and  gamma-ray range. The analytical description can be generalized to tight-focusing regime beyond the paraxial approximation considering interaction at an angle for the regions with curved wavefronts.

In summary, the findings of this manuscript are expected to be important for near-future laser-electron scattering experiments, as the calculations are fast and can provide real-time feadback during the course of an experiment.

\section{Acknowledgments}\label{sc:acknowledgments}

The authors thank B. Martinez for proofreading the manuscript, as well as L. O. Silva and T. Grismayer for fruitful discussions. This work was supported by the European Research Council (ERC-2015-AdG Grant No. 695088) and Portuguese Science Foundation (FCT) Grant No. CEECIND/01906/2018. We acknowledge PRACE for awarding access to MareNostrum based in the Barcelona Supercomputing Centre.

\section{Data availability statement}\label{sc:dataavailability}
The data that support the findings of this study are available upon reasonable request from the authors.

\vspace{10pt}
\noindent \textbf{References}
\bibliography{bibtex}

\providecommand{\newblock}{}
\begin{thebibliography}{10}
\expandafter\ifx\csname url\endcsname\relax
  \def\url#1{{\tt #1}}\fi
\expandafter\ifx\csname urlprefix\endcsname\relax\def\urlprefix{URL }\fi
\providecommand{\eprint}[2][]{\url{#2}}

\bibitem{Ritus1985}
Ritus V~I 1985 {\em Journal of Soviet Laser Research\/} {\bf 6}(5) 497617

\bibitem{Erber66}
Erber T 1966 {\em Rev. Mod. Phys.\/} {\bf 38}(4) 626--659

\bibitem{astroplasma_Uzdensky_2014}
Uzdensky D~A and Rightley S 2014 {\em Reports on Progress in Physics\/} {\bf
  77} 036902

\bibitem{Timokhin2010}
Timokhin A~N 2010 {\em Monthly Notices of the Royal Astronomical Society\/}
  {\bf 408} 2092--2114 ISSN 0035-8711 (\textit{Preprint}
  \eprint{https://academic.oup.com/mnras/article-pdf/408/4/2092/4220523/mnras0408-2092.pdf})

\bibitem{astroplasma_Medin_2010}
Medin Z and Lai D 2010 {\em Monthly Notices of the Royal Astronomical
  Society\/} {\bf 406} 1379--1404 ISSN 0035-8711 (\textit{Preprint}
  \eprint{https://academic.oup.com/mnras/article-pdf/406/2/1379/18718642/mnr0406-1379.pdf})

\bibitem{Kirk2009}
Kirk J~G, Bell A~R and Arka I 2009 {\em Plasma Physics and Controlled Fusion\/}
  {\bf 51} 085008

\bibitem{Grismayer2016}
Grismayer T, Vranic M, Martins J~L, Fonseca R~A and Silva L~O 2016 {\em Physics
  of Plasmas\/} {\bf 23} 056706 (\textit{Preprint}
  \eprint{https://doi.org/10.1063/1.4950841})

\bibitem{Luo2018}
Luo W, Liu W~Y, Yuan T, Chen M, Yu J~Y, Li F~Y, Del~Sorbo D, Ridgers C~P and
  Sheng Z~M 2018 {\em Scientific Reports\/} {\bf 8} 8400 ISSN 2045-2322

\bibitem{Elkina2011}
Elkina N~V, Fedotov A~M, Kostyukov I~Y, Legkov M~V, Narozhny N~B, Nerush E~N
  and Ruhl H 2011 {\em Phys. Rev. ST Accel. Beams\/} {\bf 14}(5) 054401

\bibitem{qu2020}
Qu K, Meuren S and Fisch N~J 2020 Signature of collective plasma effects in
  beam-driven qed cascades (\textit{Preprint} \eprint{2001.02590})

\bibitem{Bulanov2013}
Bulanov S~S, Schroeder C~B, Esarey E and Leemans W~P 2013 {\em Phys. Rev. A\/}
  {\bf 87}(6) 062110

\bibitem{Bell2008}
Bell A~R and Kirk J~G 2008 {\em Phys. Rev. Lett.\/} {\bf 101}(20) 200403

\bibitem{Bulanov2006}
Bulanov S~S, Narozhny N~B, Mur V~D and Popov V~S 2006 {\em Journal of
  Experimental and Theoretical Physics\/} {\bf 102} 9--23 ISSN 1090-6509

\bibitem{Nerush2011}
Nerush E~N, Kostyukov I~Y, Fedotov A~M, Narozhny N~B, Elkina N~V and Ruhl H
  2011 {\em Phys. Rev. Lett.\/} {\bf 106}(3) 035001

\bibitem{Mironov_2016}
Mironov A~A, Fedotov A~M and Narozhnyi N~B 2016 {\em Quantum Electronics\/}
  {\bf 46} 305--309

\bibitem{Lobet_2016}
Lobet M, d{\textquotesingle}Humi{\`{e}}res E, Grech M, Ruyer C, Davoine X and
  Gremillet L 2016 {\em Journal of Physics: Conference Series\/} {\bf 688}
  012058

\bibitem{VRANIC201565}
Vranic M, Grismayer T, Martins J, Fonseca R and Silva L 2015 {\em Computer
  Physics Communications\/} {\bf 191} 65--73 ISSN 0010-4655

\bibitem{Duclous_2010}
Duclous R, Kirk J~G and Bell A~R 2010 {\em Plasma Physics and Controlled
  Fusion\/} {\bf 53} 015009

\bibitem{Gonoskov2015}
Gonoskov A, Bastrakov S, Efimenko E, Ilderton A, Marklund M, Meyerov I,
  Muraviev A, Sergeev A, Surmin I and Wallin E 2015 {\em Phys. Rev. E\/} {\bf
  92}(2) 023305

\bibitem{Bashmakov2014}
Bashmakov V~F, Nerush E~N, Kostyukov I~Y, Fedotov A~M and Narozhny N~B 2014
  {\em Physics of Plasmas\/} {\bf 21} 013105 (\textit{Preprint}
  \eprint{https://doi.org/10.1063/1.4861863})

\bibitem{Gelfer2015}
Gelfer E~G, Mironov A~A, Fedotov A~M, Bashmakov V~F, Nerush E~N, Kostyukov I~Y
  and Narozhny N~B 2015 {\em Phys. Rev. A\/} {\bf 92}(2) 022113

\bibitem{Vranic2016_trap}
Vranic M, Grismayer T, Fonseca R~A and Silva L~O 2016 {\em Plasma Physics and
  Controlled Fusion\/} {\bf 59} 014040

\bibitem{Gonoskov2017}
Gonoskov A, Bashinov A, Bastrakov S, Efimenko E, Ilderton A, Kim A, Marklund M,
  Meyerov I, Muraviev A and Sergeev A 2017 {\em Phys. Rev. X\/} {\bf 7}(4)
  041003

\bibitem{Jirka2016}
Jirka M, Klimo O, Bulanov S~V, Esirkepov T~Z, Gelfer E, Bulanov S~S, Weber S
  and Korn G 2016 {\em Phys. Rev. E\/} {\bf 93}(2) 023207

\bibitem{SeededQEDcascades2017}
Grismayer T, Vranic M, Martins J~L, Fonseca R~A and Silva L~O 2017 {\em Phys.
  Rev. E\/} {\bf 95}(2) 023210

\bibitem{Kostyukov2016}
Kostyukov I~Y and Nerush E~N 2016 {\em Physics of Plasmas\/} {\bf 23} 093119
  (\textit{Preprint} \eprint{https://doi.org/10.1063/1.4962567})

\bibitem{Jirka2017}
Jirka M, Klimo O, Vranic M, Weber S and Korn G 2017 {\em Scientific Reports\/}
  {\bf 7} 15302 ISSN 2045-2322

\bibitem{slac1996}
Bula C, McDonald K~T, Prebys E~J, Bamber C, Boege S, Kotseroglou T, Melissinos
  A~C, Meyerhofer D~D, Ragg W and \etal B 1996 {\em Phys. Rev. Lett.\/} {\bf
  76}(17) 3116--3119

\bibitem{slac1997}
Burke D~L, Field R~C, Horton-Smith G, Spencer J~E, Walz D, Berridge S~C, Bugg
  W~M, Shmakov K, Weidemann A~W and Bula C~e 1997 {\em Phys. Rev. Lett.\/} {\bf
  79}(9) 1626--1629

\bibitem{Yakimenko2019}
Yakimenko V, Alsberg L, Bong E, Bouchard G, Clarke C, Emma C, Green S, Hast C,
  Hogan M~J, Seabury J, Lipkowitz N, O'Shea B, Storey D, White G and Yocky G
  2019 {\em Phys. Rev. Accel. Beams\/} {\bf 22}(10) 101301

\bibitem{DiPiazza2013}
Neitz N and Di~Piazza A 2013 {\em Phys. Rev. Lett.\/} {\bf 111}(5) 054802

\bibitem{Vranic2016}
Vranic M, Grismayer T, Fonseca R~A and Silva L~O 2016 {\em New Journal of
  Physics\/} {\bf 18} 073035

\bibitem{Niel2018}
Niel F, Riconda C, Amiranoff F, Duclous R and Grech M 2018 {\em Phys. Rev. E\/}
  {\bf 97}(4) 043209

\bibitem{Ridgers2017}
Ridgers C~P, Blackburn T~G, Del~Sorbo D, Bradley L~E, Slade-Lowther C, Baird
  C~D, Mangles S~P~D, McKenna P, Marklund M, Murphy C~D and et~al 2017 {\em
  Journal of Plasma Physics\/} {\bf 83} 715830502

\bibitem{Blackburn2015}
Blackburn T 2015 {\em Plasma Physics and Controlled Fusion\/} {\bf 57}

\bibitem{tamburini2020}
Tamburini M 2020 On-shot diagnostic of electron beam-laser pulse interaction
  based on stochastic quantum radiation reaction (\textit{Preprint}
  \eprint{2007.02841})

\bibitem{Blackburn2020ind}
Blackburn T~G, Gerstmayr E, Mangles S~P~D and Marklund M 2020 {\em Phys. Rev.
  Accel. Beams\/} {\bf 23}(6) 064001

\bibitem{Cole2018}
Cole J~M, Behm K~T, Gerstmayr E, Blackburn T~G, Wood J~C, Baird C~D, Duff M~J,
  Harvey C, Ilderton A and Joglekar A~S~e 2018 {\em Phys. Rev. X\/} {\bf 8}(1)
  011020

\bibitem{Poder2018}
Poder K, Tamburini M, Sarri G, Di~Piazza A, Kuschel S, Baird C~D, Behm K,
  Bohlen S, Cole J~M and Corvan D~J~e 2018 {\em Phys. Rev. X\/} {\bf 8}(3)
  031004

\bibitem{ELI}
Eli science and technology with ultra-intense lasers whitebook, edited by
  andreas thoss (2011)

\bibitem{Apollon}
Apollon \urlprefix\url{http://www.apollon-laser.fr/}

\bibitem{CoReLS}
Corels \urlprefix\url{https://corels.ibs.re.kr/html/corels_en/}

\bibitem{FACET-II}
Facet-ii
  \urlprefix\url{https://portal.slac.stanford.edu/sites/ard_public/facet/Pages/
  FACET-II.aspx}

\bibitem{meuren2020seminal}
Meuren S, Bucksbaum P~H, Fisch N~J, Fiúza F, Glenzer S, Hogan M~J, Qu K, Reis
  D~A, White G and Yakimenko V 2020 On seminal hedp research opportunities
  enabled by colocating multi-petawatt laser with high-density electron beams
  (\textit{Preprint} \eprint{2002.10051})

\bibitem{abramowicz2019letter}
Abramowicz H, Altarelli M, Aßmann R, Behnke T, Benhammou Y, Borysov O,
  Borysova M, Brinkmann R, Burkart F and \etal K~B 2019 Letter of intent for
  the luxe experiment (\textit{Preprint} \eprint{1909.00860})

\bibitem{EXCELS}
Excels \urlprefix\url{http://www.xcels.iapras.ru.}

\bibitem{ZEUS}
Zeus \urlprefix\url{https://zeus.engin.umich.edu}

\bibitem{breitwheeler1934}
Breit G and Wheeler J~A 1934 {\em Phys. Rev.\/} {\bf 46}(12) 1087--1091

\bibitem{Blackburn2017}
Blackburn T~G, Ilderton A, Murphy C~D and Marklund M 2017 {\em Phys. Rev. A\/}
  {\bf 96}(2) 022128

\bibitem{Fonseca2002}
Fonseca R~A, Silva L~O, Tsung F~S, Decyk V~K, Lu W, Ren C, Mori W~B, Deng S,
  Lee S and Katsouleas T~e 2002 Osiris: A three-dimensional, fully relativistic
  particle in cell code for modeling plasma based accelerators {\em
  Computational Science --- ICCS 2002\/} ed Sloot P~M~A, Hoekstra A~G, Tan
  C~J~K and Dongarra J~J (Berlin, Heidelberg: Springer Berlin Heidelberg) pp
  342--351 ISBN 978-3-540-47789-1

\bibitem{abramowicz2021conceptual}
Abramowicz H, Acosta U~H, Altarelli M, Assmann R, Bai Z, Behnke T, Benhammou Y,
  Blackburn T, Boogert S and \etal O~B 2021 Conceptual design report for the
  luxe experiment (\textit{Preprint} \eprint{2102.02032})

\bibitem{DiPiazzaStochastic}
Neitz N and Di~Piazza A 2013 {\em Phys. Rev. Lett.\/} {\bf 111}(5) 054802

\bibitem{GreenHarvey2014}
Green D~G and Harvey C~N 2014 {\em Phys. Rev. Lett.\/} {\bf 112}(16) 164801

\bibitem{Yan2017}
Yan W, Fruhling C, Golovin G, Haden D, Luo J, Zhang P, Zhao B, Zhang J, Liu C,
  Chen M, Chen S, Banerjee S and Umstadter D 2017 {\em Nature Photonics\/} {\bf
  11} 514--520 ISSN 1749-4893

\bibitem{Mackenroth2019}
Mackenroth F, Holkundkar A~R and Schlenvoigt H~P 2019 {\em New Journal of
  Physics\/} {\bf 21} 123028

\bibitem{Hu2020}
Hu G, Sun W~Q, Li B~J, Li Y~F, Wang W~M, Zhu M, Hu H~S and Li Y~T 2020 {\em
  Phys. Rev. A\/} {\bf 102}(4) 042218

\end{thebibliography}

\end{document}